\documentclass[structabstract]{aa}
\usepackage{txfonts}

\usepackage[authoryear]{natbib}
\usepackage{graphicx}
\usepackage{txfonts}

\def\degr{\hbox{$^\circ$}}
\def\arcmin{\hbox{$^\prime$}}
\def\arcsec{\hbox{$^{\prime\prime}$}}

\begin{document}

\title{Spatial variation of the cooling lines in the Orion Bar from
  Herschel/PACS} 


\author{J. Bernard-Salas\inst{1} 
\and E. Habart \inst{1} 
\and H.Arab\inst{1}
\and A. Abergel\inst{1} 
\and E. Dartois\inst{1} 
\and P. Martin\inst{2}
\and S. Bontemps\inst{3} 
\and C. Joblin\inst{4,5} 
\and G.J. White\inst{6,7} 
\and J.-P. Bernard\inst{4,5} 
\and D. Naylor\inst{8}}

\offprints{J. Bernard-Salas, \email{jbernard@ias.u-psud.fr}}

\institute{Institut d'Astrophysique Spatiale, Paris-Sud 11, 91405
  Orsay, France 
\and Canadian Institute for Theoretical Astrophysics, Toronto, Ontario,
  M5S 3H8, Canada 
\and CNRS/INSU, Laboratoire d'Astrophysique de Bordeaux, UMR 5804, BP 89, 33271
  Floirac Cedex, France 
\and Universit\'e de Toulouse, UPS-OMP, IRAP, Toulouse,
  France
\and CNRS, IRAP, 9 Av. colonel Roche, BP 44346, 31028 Toulouse Cedex 4,
  France
\and Dept. of Physicics \& Astronomy, The Open University, Milton
  Keynes MK7 6AA, UK
\and Space Science \& Technology Division, The Rutherford Appleton
  Laboratory, Chilton, Didcot OX11 0NL, UK
\and Institute for Space Imaging Science, University of Lethbridge,
  Lethbridge, Canada}

\date{Received date / Accepted date}

\abstract {The energetics in photo-dissociation regions are
  mainly regulated by the balance between the heating from the
  photo-electric effect acting on dust grains, and the cooling via the
  copious emission of photons in far-infrared lines.
The Orion Bar is a
  luminous and nearby photo-dissociation region,
  which presents to the observer an ideal edge-on orientation in which
  to study this energy balance. 
Spatially resolved studies of such a nearby system are essential as they enable us to
  characterise the physical processes that control the energetics of the
  regions and can serve as templates for distant systems where
  these processes cannot be disentangled.
}
{To characterise the emission of the far-infrared fine-structure lines of
  \ion{[C}{II]} (158$\mu$m), \ion{[O}{I]} (63 and 145$\mu$m), and
\ion{[N}{II]} (122$\mu$m) that trace the gas local conditions, via
spatially resolved observations of the Orion Bar. The observed
distribution and variation of the lines are discussed in relation to
the underlying geometry and linked to the energetics associated with the Trapezium stars.} 
{Herschel/PACS observations are used to map the spatial distribution
  of these fine-structure lines across the Bar, with a  spatial
  resolution between 4\arcsec~and 11\arcsec~and covering a total square area of
  about 120$\arcsec$x105$\arcsec$. The spatial profile of the emission
  lines are
  modelled using the radiative transfer code Cloudy.}
{The Herschel observations reveal in unprecedented detail the morphology of the
  Bar.
 The spatial distribution of the \ion{[C}{II]} line coincides with
  that of the \ion{[O}{I]} lines. The \ion{[N}{II]} line peaks closer to the
  ionising star than the other three lines, but with a small region of
  overlap. We can distinguish  several knots of enhanced
  emission within the Bar indicating the presence of an inhomogenous
  and structured 
  medium.
  The emission profiles cannot be reproduced by a single
  photo-dissociation region, clearly indicating that, besides the Bar,
  there is a significant contribution from additional photo-dissociation
  region(s) over the area studied.
 The combination of both the \ion{[N}{II]} and \ion{[O}{I]} 145$\mu$m
  lines can be used to estimate the \ion{[C}{II]} emission and
  distinguish between its ionised or neutral origin.
 We have calculated how much \ion{[C}{II]} emission comes
  from the neutral and ionised region, and find that at least
    $\sim$82\% originates from
  the photo-dissocciation region. Together,  the \ion{[C}{II]} 158$\mu$m and
  \ion{[O}{I]} 63 and 145$\mu$m lines account for $\sim$90\% of the
  power emitted by the main cooling lines in the Bar (including CO, H$_2$,
  etc...), with \ion{[O}{I]} 63$\mu$m alone accounting for 72\% of the total.}  {}


\authorrunning{Bernard-Salas et al.}
\titlerunning{Cooling lines in the Orion Bar}  

\maketitle

\section{Introduction}

Most of the mass in the Interstellar Medium (ISM) is in neutral form.
Photo-dissociation regions (PDRs) are neutral regions of the ISM at
the interface between the stars and opaque cores of molecular clouds, where the
heating and chemistry are regulated by the penetrating far ultraviolet
(FUV) photons of the ionising source(s) \citep[e.g.][]{hol99, ber05}. The gas
in the PDRs cools mainly via the emission of far-infrared (FIR) fine
structure lines such as \ion{[C}{ii]}, \ion{[O}{i]}, and molecules
(e.g. CO, H$_2$).  The intensity of these lines depends on local
conditions in the gas, and when compared to PDR models can be used
to trace the spatial evolution of the physical conditions of the gas
across the PDRs.

Due to its proximity \citep[415pc,][]{men07} and edge-on orientation,
the Orion Bar has been subject to extensive studies in the literature,
where it is usually adopted as the prototypical PDR template in the
study of high mass star forming regions, and is often used to test PDR
models \citep[e.g.,][]{pel09}. It is one of the brightest PDRs with an
FUV radiation field at the ionisation front of  G$=$1-4$\times$10$^{4}$ G$_{\rm 0}$ \citep{tie85,mar98}, where  
G$_{\rm 0}$ is the mean interstellar radiation field \citep[1.6$\times$10$^{-6}$ W~m$^{-2}$,][]{hab68}. 
 \citet{tie93} showed how the Bar presented
a layered distribution of Polycyclic Aromatic Hydrocarbons (PAHs),
H$_2$, and CO. This distribution is thought to be the result of an
extended gas component seen edge-on, of average gas density between
10$^4$-10$^5$ cm$^{-3}$. Evidence of clumps have also been suggested
to explain high density and temperature tracers
\citep[e.g.,][]{bur90,vdw96}. Recently, \citet{rub11} presented mid-IR
Spitzer observations that extend beyond the Bar, from 2.1 to up to
12\arcmin.1 from the ionising star $\theta$$^1$ Ori C. They detect
ionised material (Ne$^{+2}$, IP of 41eV) all the way up to the
boundary of their observations. They also find evidence for a decrease
in electron density and an increase of the PDR tracers ([Si II] 34.8
$\mu$m, [Fe II] 26.0 $\mu$m, and molecular hydrogen) as the distance
from $\theta$$^1$ Ori C increases.  In the FIR, \citet{her97} mapped
the Orion region with a 22$\arcsec$ to 55$\arcsec$ beam using the Kuiper Airborne
Observatory (KAO). Their derived column densities in the Bar for O$^0$
and C$^+$ were in agreement with an edge-on geometry.

In this paper, we present the first maps of the \ion{[C}{II]}
(158$\mu$m), \ion{[O}{I]} (63 and 145$\mu$m), and \ion{[N}{II]}
(122$\mu$m) lines of the Orion Bar region from the recently launched
Herschel Space Observatory \citep[HSO,][]{pil10}.  With its access to
FIR and superb spatial resolution at these wavelengths, the HSO allows
us to study the spatial distribution of these  lines in
unprecedented detail and so improve our understanding of the ionised and
neutral interface. These observations enable us to trace the spatial
evolution and excitation conditions of the gas across the illuminated
interface of the Bar.  This paper is part of a Herschel study of the
Orion Bar by our group and is complemented by Herschel/SPIRE studies of the CO
and CH$^+$ line emission \citep{hab10,hab11,nay10}, and dust emission \citep{ara11}.

The paper is organized as follows. The observations, data reduction
and line measurements are given in Sect.~2. In the following
section (Sect.~3) the spatial morphology of the lines is discussed. In Sect.~4
the distribution and correlation of the lines are analysed. Sect.~5
presents a model tailored to reproduce the distribution of the
lines. Discussions on the origin of the \ion{[C}{II]} emission and
the cooling are presented in Sect.~6 and Sect.~7. 
Finally, the conclusions are summarized in the last section.

\section{Observations and Data Reduction}

\begin{figure}[!t]
  \begin{center}
   \includegraphics[width=8cm]{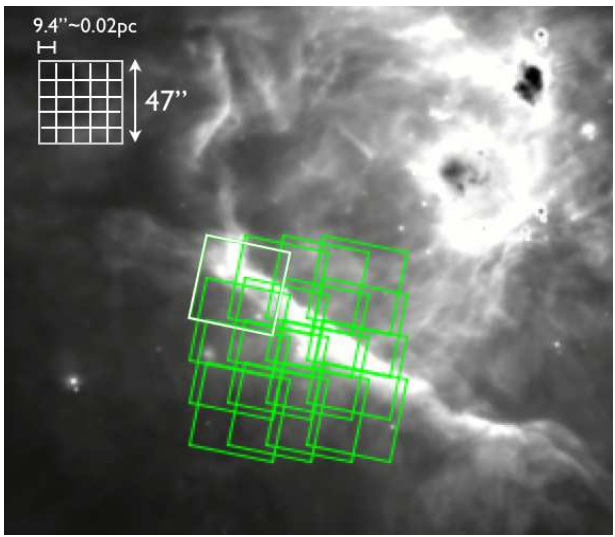}
  \end{center}
  \caption{Overlay of the \ion{[O}{I]} 63$\mu$m PACS observation on a
    Spitzer/IRAC 8$\mu$m image of the Orion Bar. A 4$\times$4 raster
    map - 16 overlapping footprints - was performed (see Sect. 2). An
    example of a footprint with its 5$\times$5 {\em spaxels} is
    illustrated on the top left of the figure together with their
    dimensions. The map covers the region before and behind the
    Bar (toward $\theta$$^1$ Ori C).}
\end{figure}

The observations were taken using the PACS instrument \citep{pog10} on
board the HSO on 24 February 2010. These observations are part of the
{\em Evolution of the Interstellar Medium} guaranteed time key project
(observation ID=1342191152) \citep{abe10}. Four fine-structure lines
were targeted: \ion{[C}{II]} at 158$\mu$m, \ion{[O}{I]} at 63 and
145$\mu$m, and \ion{[N}{II]} at 122$\mu$m. At these wavelengths,
  and in increasing order, PACS provides a spectral resolution of
  3420, 1050, 1170, and 1270 respectively (v$\sim$90-285
  km~s$^{-1}$). The observations were taken during Science
Demonstration Phase (SDP) in the - now decommissioned -
wavelength switching mode.

To trace the PDR and H\,II region interface a 4x4 raster map was
performed. Each raster position (or footprint) is composed by
5$\times$5 spatial pixels referred to as {\em spaxels}. For each
spaxel the line is observed in 16 different spectral scans, each with
an up and down scan.  The configuration at the time of the observation
is shown in Figure~1, where the raster map at the epoch of observation
is overlaid on top of an 8~$\mu$m IRAC image of the Orion Bar. We note
that an additional map in the direction of the Trapezium stars and
overlapping the current observation is currently scheduled for
observation and will be presented in a future paper. Given the
brightness of the lines, the minimum exposure configuration of one
cycle and repetition per line was performed.  For these maps the best
sampling is achieved using Nyquist sampling, which we adopted for the
lines measured in the red channel (\ion{[C}{II]} at 158$\mu$m,
\ion{[O}{I]} 145$\mu$m, and \ion{[N}{II]} at 122$\mu$m). This consists
of a raster point and raster line step
of 24\arcsec.0 and 22\arcsec.0 respectively. This results approximately in steps of
2/3 and 1/2 of the slit size along both directions.  For the
\ion{[O}{I]} 63$\mu$m line located in the red channel, Nyquist
sampling is achieved with 16\arcsec.0~ and 14\arcsec.5 raster point
and line steps. However, to minimise exposure time, and given the
strength of the line, the same step size as the lines for the red
channel was adopted. An {\em off} position, about 40\arcmin~from the
center of the map, was taken at the beginning and the end of the
observation (not shown in the figure).  This observation is not
subtracted from the data as the purpose of the wavelength switching
mode is to cancel out the background by determining a differential
line profile. The {\em off} observation, however, confirmed that there
 was
no background contamination to any of the lines studied.  
 
The data were processed using version 6.1.0 of the reduction and
analysis package HIPE \citep{ott10}. From level 0.5 to level 1 the
standard procedures were followed. From this level on the cubes were
further processed, using proprietary tools, to  correct for
drifting and for flux misalignments between scans. Drifting is caused
by temperature deviations in the telescope which can cause the signal
to be modified over time. They can be easily identified by comparing
the up and down scans. To correct for this effect we first passed a
median filter in the time domain to avoid glitches. The continuum was
then fitted using a linear fit\footnote{In our data the signal was
  either linear or slightly curved. A polynomial fit of second degree
  produced very similar results.} to characterise and remove the
drift. The flux misalignment between the spectral scans can be due to
improper dark subtraction or flatfield correction. The former will
have an additive effect, and the latter a multiplicative one. In our
case, the scans could be well aligned using an additive value which
indicates that in our observations the cause of the flux misalignments
is the dark subtraction. We note that the lines are very strong and
these corrections were minor.

At this point the cubes are exported and we used the IDL-based
software PACSman \citep{leb11} to measure the line fluxes (by fitting
a Gaussian) and create the final map. In this code, the line fluxes
are measured for all the spaxels independently. The
lines are not spectrally resolved.  To create the final
map, PACSman recreates an oversampled pixelated grid of the
observations (with 3$\arcsec$ pixel resolution) and calculates the
average fractional contribution of the given spaxels to the relevant
position. The code calculates the statistical uncertainties on the
fly, which include the dispersion in the reduction process and the RMS
of the fit. These uncertainties are small and usually amount to less
than 4\%, 1\%, 3\% for the \ion{[C}{II]} and \ion{[O}{I]} lines.  For
the weaker \ion{[N}{II]} 122$\mu$m line, these are higher and
oscillate between 5-13\%. The relative accuracy between spaxels given
in the manual
is 10\%\footnote{From the PACS spectroscopy performance and
  calibration manual. This can be found at
  http://herschel.esac.esa.int/twiki/bin/view/Public/PacsCalibrationWeb}, and for the remaining of the paper we have adopted the higher
of this value or the uncertainty given by PACSman. We note that
absolute flux calibration is quoted to be 30\% of the peak-to-peak
accuracy (scatter around the expected flux densities).  The final maps
for each of the four lines are shown in Figure~2.  We have compared
our \ion{[C}{II]} and \ion{[O}{I]} line fluxes with those measured by
\cite{her97} with the Fabry-P\'erot interferometer FIFI and integrated
over a 55\arcsec~beam. Their measurements are in good agreement
(within $\sim$30\%) with those reported in this paper when convolved
to their larger beam size.

\label{obs_s}

\begin{figure*}[!ht]
  \begin{center}
  \includegraphics[width=18cm]{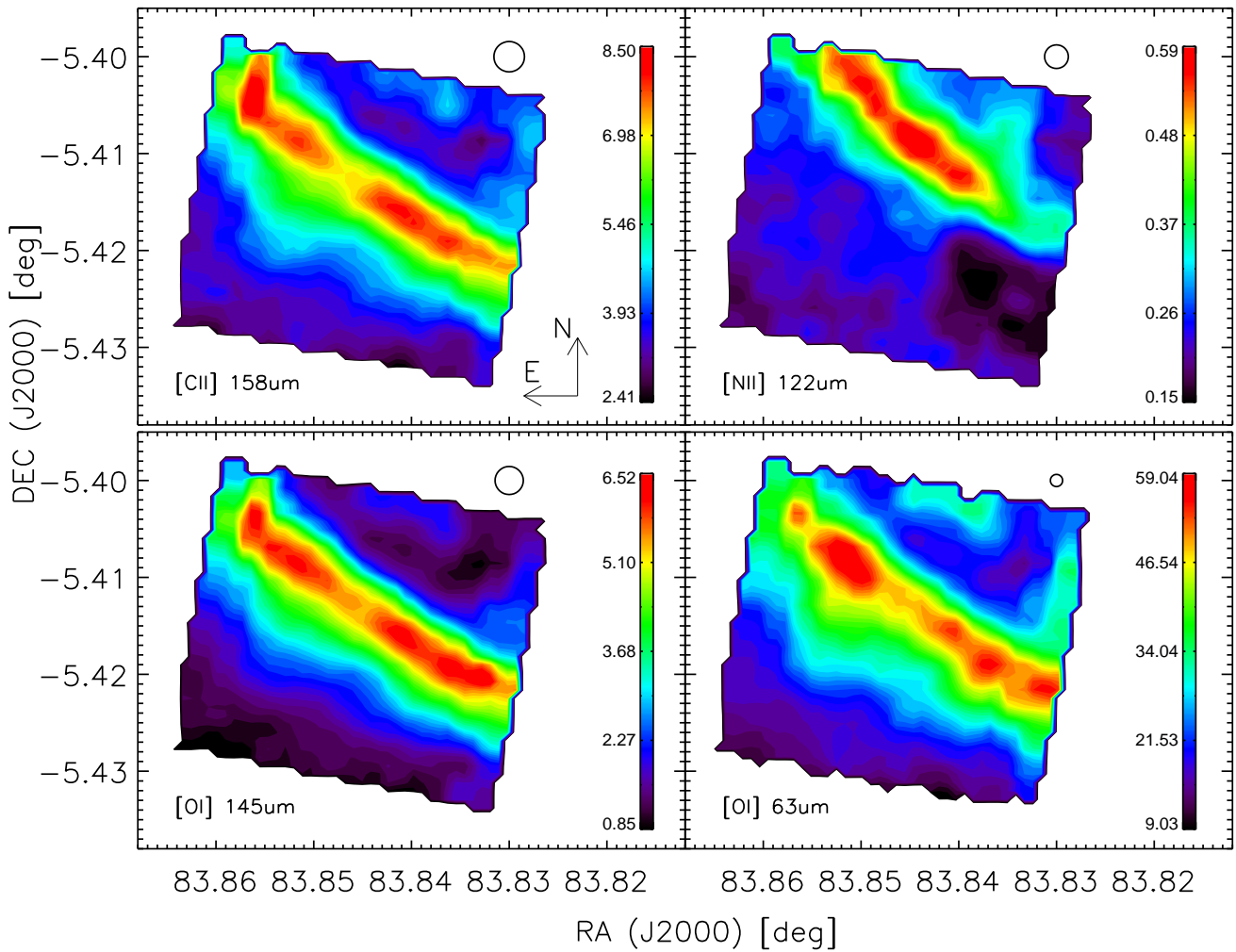}
  \end{center}
  \caption{PACS contour maps of the \ion{[C}{II]} (158$\mu$m),
    \ion{[O}{I]} (63 and 145$\mu$m), and \ion{[N}{II]} (122$\mu$m)
    lines in units of 10$^{-6}$ W~m$^{-2}$~sr$^{-1}$. In increasing
    order of wavelength the beams sizes are 4\arcsec.5, 8\arcsec.8,
    10\arcsec, and 11\arcsec. Note the offset from zero in the
    intensity scale to emphasize the detail.}
\end{figure*}

\begin{figure*}
  \begin{center}
 \includegraphics[width=16.cm]{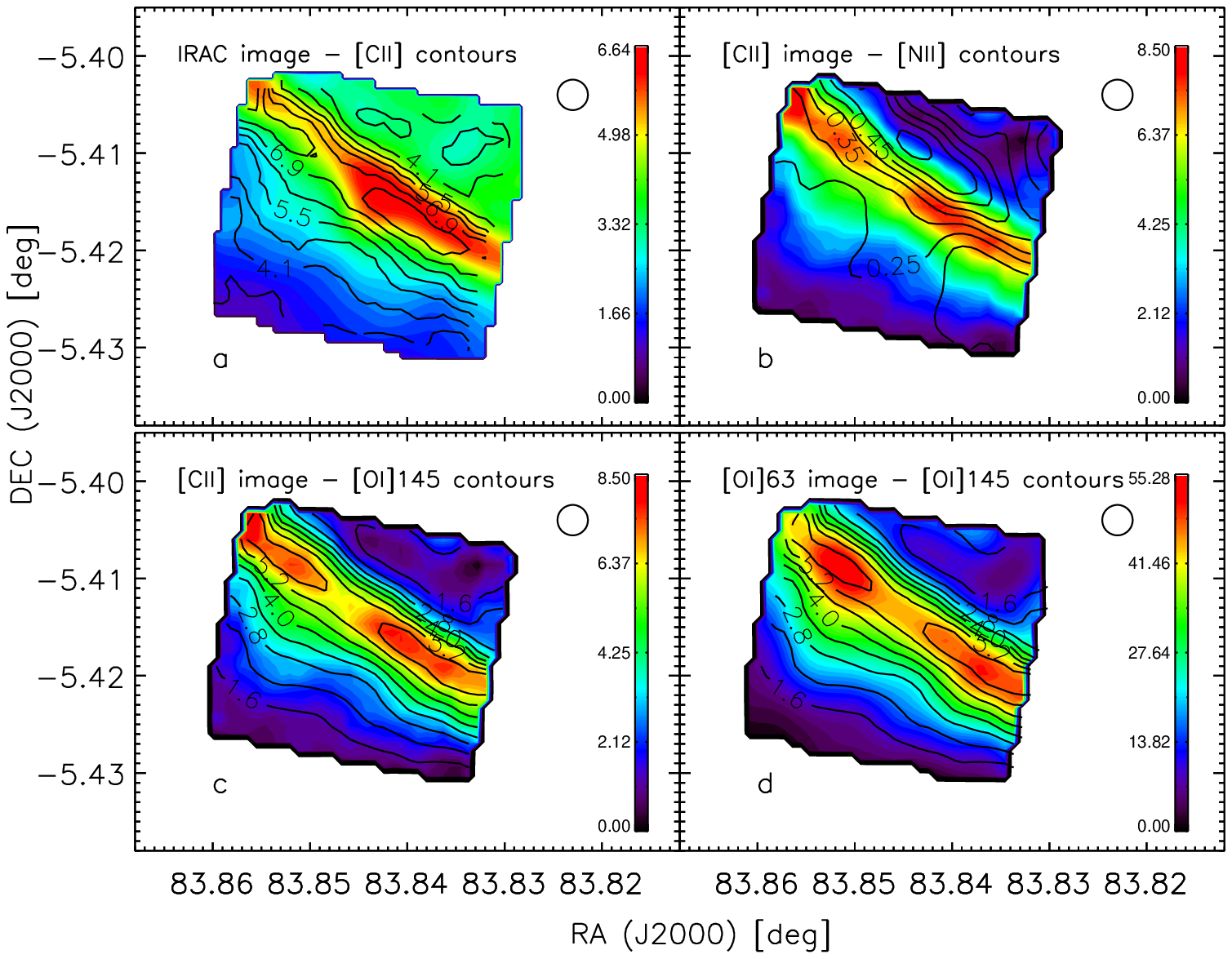}
  \end{center}
  \caption{Image and contour plots for different combination of lines
    and 8$\mu$m band as indicated in the title of each panel. All maps have been
    convolved to the PSF at 158$\mu$m and are in flux units of
    10$^{-6}$ W~m$^{-2}$~sr$^{-1}$ (except for the IRAC image which is
    in 10$^{3}$MJy~sr$^{-1}$. Here the intensity scales start at zero.}
\end{figure*}

\section{Spatial Distribution}

The observed spatial distribution of the lines is shown in Figure~2,
where the ionising star $\theta^1$ Ori C is illuminating the region
from the top right, outside the maps (see also Fig.4a for a sketch of
the different regions). The beams sizes are represented in the top
right of each map and correspond, in increasing order of wavelength,
to 4\arcsec.5, 8\arcsec.8, 10\arcsec, and 11\arcsec. At the distance
of the Bar (415~pc), 10$\arcsec$ corresponds to a physical scale of
0.02~pc. The Bar\footnote{For the remainder of the paper, the Bar is
  defined as the region where the emission is higher than 75\% of the
  peak emission for carbon. This threshold falls between the third and
  fourth highest  contours in  Figure~3 and is represented by the
  solid line in Figure~4a. Similarly, we use the same  definition for
  the discussion in Sect. 6 about the \ion{[N}{II]} peak.}  is resolved in
all four lines and we detect emission all over the
region probed, including detection of \ion{[N}{ii]} (ionised gas)
behind the Bar. The range of intensities is not large, with variations
by factors 4 (\ion{[C}{II]} and \ion{[N}{II]}) to 8 (\ion{[O}{I]}) from the peak (Bar) to the fainter (outer)
regions.

The overall morphology of the region can be best discussed from the
\ion{[O}{I]} 63$\mu$m map because its smaller point spread function
(PSF) provides the highest detail (Figure~2).  Even if this line is
more affected by self-absorption, the gain in resolution bests this
caveat.  The emission peaks at about 123$\arcsec$ SE of $\theta^1$ Ori
C. The most striking feature is the presence of several knots of
enhanced emission resulting in small scale structures, which
suggests a clumpy distribution within the Bar. These knots, three in
the south-west and the two in the north-east, are bridged by weaker
emission, $\sim$16\% lower in flux than the knots. In the
  \ion{[O}{I]} 63$\mu$m image, the smaller knots have diameters
  between $\sim$6$\arcsec$ and $\sim$10$\arcsec$ which, at the distance of the Orion nebula,
  correspond to 0.01-0.02 pc. The smaller ones (6\arcsec) are marginally resolved and could result from the superposition of even smaller clumps.  These knots could be clumps being photo-evaporated by the intense FUV penetrating
  the PDR. \citet{gor02} studied the effects of the radiation field on
  the evolution of clumps in PDRs. Using their relation between the size of the clumps and the column density of the FUV 
  heated region (N$_0$=2$\times$10$^{21}$ cm$^{-2}$, their  Eq.~34), we derive a density of the gas at the base of the
   photo-evaporating flow of 6-10$\times$10$^{4}$ cm$^{-3}$. This value is consistent with the inter-clump density we use in Sect. 5 for the modelling.  Knowing that the density inside the clumps is higher 
   than this, and assuming that the observed size is that of an isolated clump, we estimate that the photo-evaporation timescale of these 
   clumps should be higher than $\sim$4\,500-7\,500 yr, depending on size.

This morphology is somewhat mimicked in the \ion{[O}{I]} 145$\mu$m and
\ion{[C}{II]} 158$\mu$m maps, where the lower resolution has washed
out the clumps of emission. As we move away from the Bar (both in
front and behind) the emission decreases gradually. There is, however,
a confined region of brighter emission in the north of the map (middle
top of the \ion{[O}{I]} 63$\mu$m map), probably due to an increase in
density  in the PDR which lies behind the HII
  region. Alternatively this could be the result of a geometrical
  effect where the inclination of the background PDR is steeper. This
region is also revealed in the other three maps, albeit with less
contrast. The \ion{[O}{I]} 63$\mu$m map also points to an increased
emission at the western edge of the Bar and extending north. This
excess was also detected in CO emission from the ground
\citep[e.g.,][]{lis98}.

The morphology of the \ion{[C}{II]} line follows very well that of the
\ion{[O}{I]} lines (especially that of \ion{[O}{I]} 145$\mu$m) with
the Bar peaking at the same position in the three maps. This is better
seen in Figures~3c and 3d where the images have been
convolved to the same resolution as the
\ion{[C}{II]}158$\mu$m map (largest PSF), and are shown with
over-plotted contours. In Figure~3a we also see that
the PAH emission, as traced by the 8$\mu$m IRAC band, is close to the
\ion{[C}{II]} peak but slightly shifted towards the ionising star $\theta$$^1$ Ori C.
The peak of \ion{[C}{ii]} emission in the top-left corner of the Bar
is, however, not followed as well by the PAH emission.  In 
Figure~3d we can see that the different intensity levels
in the 145$\mu$m map delineate the Bar very well, whereas in the
63$\mu$m map the spread in the levels is broader.  This is probably
the result of the higher optical depth on the 63$\mu$m line, where
changes in column density have a less pronounced effect than for the 145$\mu$m
line.

The peak of the \ion{[N}{II]} emission is displaced with respect to the
other lines by about 12$\arcsec$~towards to $\theta^1$ Ori C (Fig. 2
and Fig. 3b). This is expected because with an ionisation potential of
14.5~eV the \ion{[N}{II]} line comes from the ionised phase and is
representative of the ionisation front. There is a small region of
overlap between the \ion{[N}{II]} and \ion{[C}{II]} lines but this
could be due to an orientation effect (Bar slightly tilted).  We even
note that the north-east tip (top-left) of the Bar that is seen in the
\ion{[C}{II]} 158$\mu$m and \ion{[O}{I]} 145$\mu$m maps (and also
8$\mu$m IRAC map, Figs. 1 and 2), delineates the boundary of emission
of the \ion{[N}{II]} line in that region. On the western side of the
Bar, the \ion{[N}{II]} emission is somewhat delineated by the
above-mentioned emission extending north in the \ion{[O}{I]} 63$\mu$m
map and also detected in CO and dust emission.  Finally, the gap in
the center of the Bar, seen in the \ion{[O}{I]} and \ion{[C}{II]}
lines, coincides with the maximum emission of the \ion{[N}{II]}
122$\mu$m line in front of it (Fig. 3b). This could indicate that
the radiation field in this specific spot is higher (or more intense)
and has been able to photo-ionise the gas. Alternatively this could be
due to variations in the column density.

\section{Line Correlations and Ratios}

\begin{figure*}
  \begin{center}
  \includegraphics[width=9.3cm,angle=90]{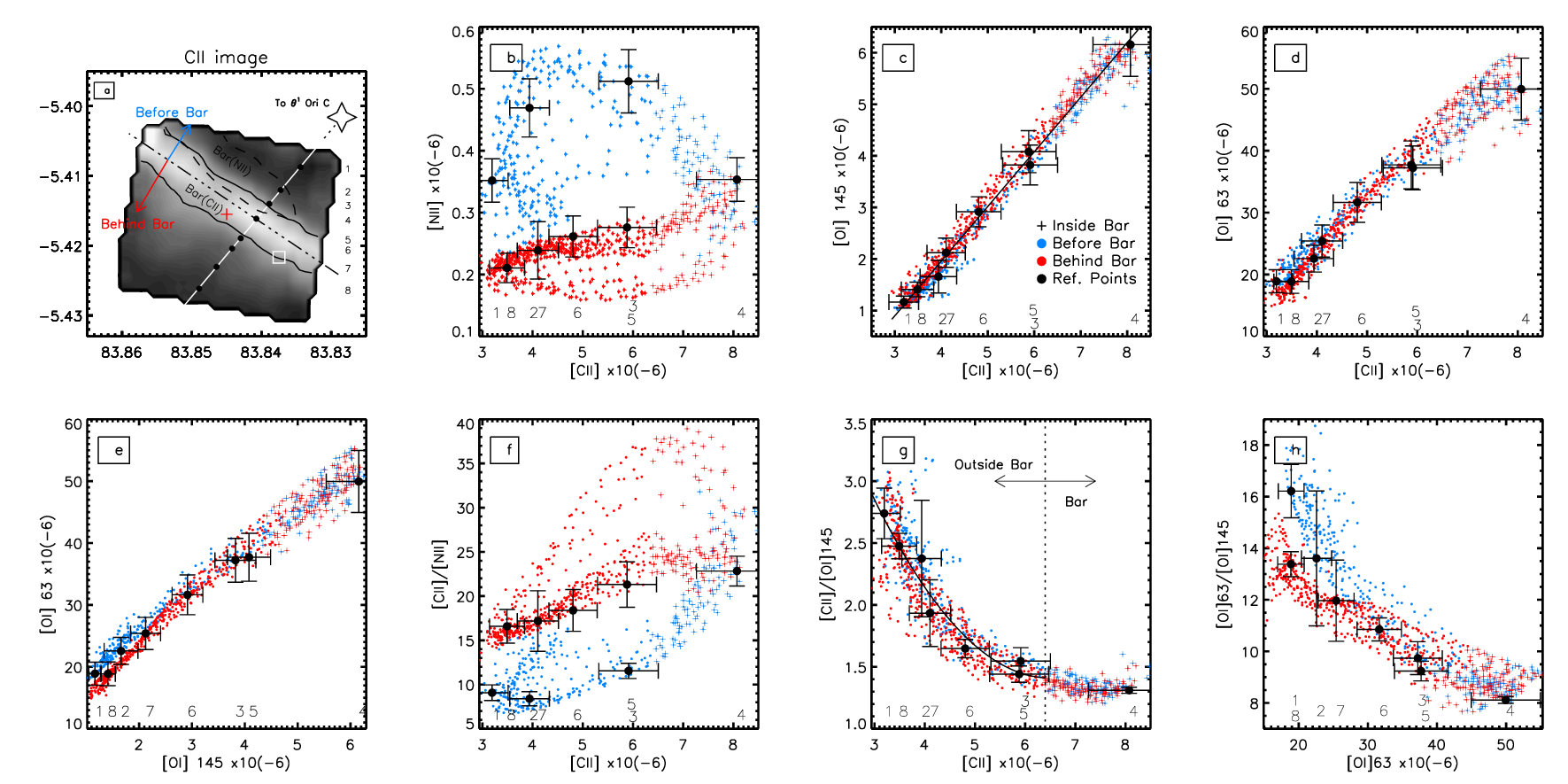}
  \end{center}
  \caption{Intensity plots for different combination of lines fluxes
    (in 10$^{-6}$ W~m$^{-2}$~sr$^{-1}$) and ratios from the convolved
    maps. The different regions are indicated in the first panel and
    are labelled in the third, where blue represents points in front
    of the Bar and red behind the Bar. Plus symbols indicate points
    within the Bar(CII). The square data point is the position where
    the cooling is calculated (see Sect. 7). The reference points for the adopted cut
    (Sect. 5) are plotted in circles and include error bars to give an
    indication of the uncertainties at different regions. In panels
    {\em c} and {\em g} the solid lines represent respectively a
    linear fit ($y=-2.28+1.06x$), and a polynomial fit for the fainter
    region outside the Bar ($y=6.37-1.53x+0.12x^2$).}
\end{figure*}

\begin{figure*}
  \begin{center}
   \includegraphics[width=16cm]{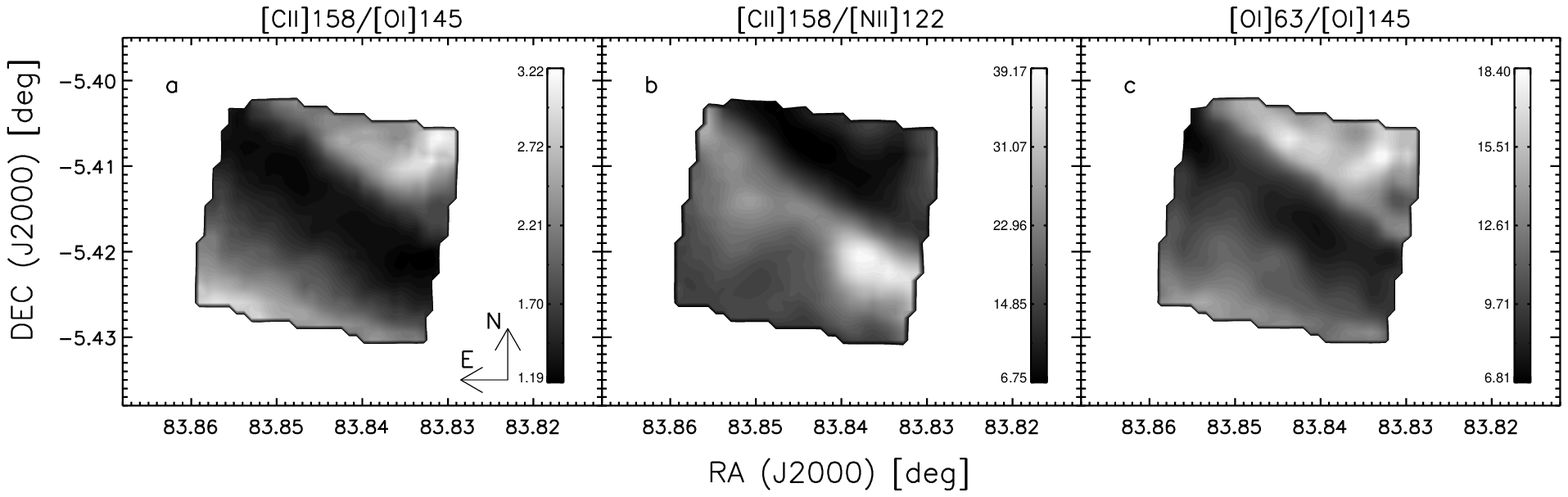}
  \end{center}
  \caption{Map ratios for different combination of lines as indicated
    in the title of each panel.}
\end{figure*}

For the purpose of the discussion we have divided the map into several regions which are represented in  Figure~4a 
and annotated  in panel {\em c} of the same figure. Summarizing: data points in front
of and behind the Bar are colour-coded in blue and red, respectively; plus
symbols indicate points that trace the Bar in \ion{[C}{II]}
(those inside the solid contour line), and dots points outside the
Bar in \ion{[C}{II]}. In addition, selected points along an adopted cut (see \S5) have been labeled (orange) and include error bars.


Figure~4b plots the \ion{[N}{II]} versus the \ion{[C}{II]} line
fluxes. The points before and behind the Bar (blue and red
respectively) occupy different regions but there are no clear trends.
\citet{abe05} hinted, from their models, at a correlation between
these two lines in the ionised medium. The intricate behaviour in our
figure does not suggest that. On the other hand, Figure~4c illustrates
that the \ion{[C}{II]} 158$\mu$m line follows very well the
\ion{[O}{I]} 145$\mu$m line. A linear fit gives a slope of
1.059$\pm$0.005. Variation of the \ion{[O}{I]} 145$\mu$m line is
dominated by the column density of the gas, and so this correlation
indicates that most of the \ion{[C}{II]} comes from the PDR and not
the HII region (see Sect. 6).

The \ion{[O}{I]} 63$\mu$m and
\ion{[C}{II]} 158$\mu$m lines show also a good correlation
(Fig.~4d) but it seems that the correlation outside the Bar(CII)
differs with respect to the other regions. This is also seen in
Figure~4e where the two oxygen lines are plotted against each
other. The points behind the Bar (red dots) are displaced compared to
those before the Bar (blue dots), and there is a change of slope between
points in the Bar (plus signs) and outside the Bar (dots). The
excitation conditions of these lines are similar and therefore this behaviour
could reflect the effect of opacities that more strongly affect the
\ion{[O}{I]}63$\mu$m line (as it becomes self-absorbed at rather low
column densities).

In the following panels we plot the ratio of several lines which, 
  in
an optically thin environment,
allows us to remove the effect of column density in the trends:

\begin{itemize} 

\item[$\bullet$] Figure~4f shows that the
  \ion{[C}{II]}(158$\mu$m)/\ion{[N}{II]}(122$\mu$m) ratio is different
  before and behind the Bar. A map of this line ratio is
  illustrated Figure~5b where we clearly distinguish two regions, with
  the points before the Bar, which are more representative of the
  ionised gas, having a lower ratio. We find that a threshold of 15
  for this ratio could be used to distinguish between emission from
  the neutral and ionised region.

\item[$\bullet$] The \ion{[C}{II]}158$\mu$m/\ion{[O}{I]}145$\mu$m
  ratio in Figure~4g decreases as we approach the Bar. The trend in
  the figure of the points outside the Bar (dots) can be approximated
  by a polynomial fit. This behaviour is again evident in the map
  ratio shown in Figure~5a where points before and behind the Bar show
  the similar values. This suggests similar physical conditions in
  these two regions because the excitation parameters for both lines
  are different.  This is not expected for a single PDR and suggests contamination of a background PDR (see Sect. 5) Another important point is that within the Bar the
  ratio is very homogeneous (within 16\%) but it differs from the
  regions outside the Bar.  We can use the
  \ion{[C}{II]}158$\mu$m/\ion{[O}{I]}145$\mu$m ratio to put a
  constraint on the densities by comparing with the predicted values
  from PDR models \citep{fer98,kau99}. We find densities that vary
  from 10$^4$-10$^5$cm$^{-3}$ in the Bar, and 10$^3$-10$^4$cm$^{-3}$
  outside it.

\item[$\bullet$] In the last panel Figure~4h, it is clear that the
  line ratio \ion{[O}{I]}63$\mu$m/\ion{[O}{I]}145$\mu$m splits before
  and behind the Bar (being lower before the Bar). Given that the
  excitation conditions in these regions must be similar (Fig. 4g), we
  conclude that the observed difference is the result of \ion{[O}{I]}
  63$\mu$m becoming self-absorbed; in the Bar, as it is expected, the
  ratio is smaller because the column density is higher than outside
  the Bar. We note that the gas in the region where the
    \ion{[O}{I]}63$\mu$m line is optically thick must still be warm
    to excite the \ion{[O}{I]}145$\mu$m line (E$_u=$326~K). From Figure~5c
  we see that the ratio is indeed not homogeneous within the Bar (as
  it was for the \ion{[C}{II]} and \ion{[O}{I]} 145$\mu$m, Fig.~5a),
  indicating that opacity effects are present.

\end{itemize}

\section{Modelling}

\begin{figure}
  \begin{center}
   \includegraphics[width=9cm,angle=90]{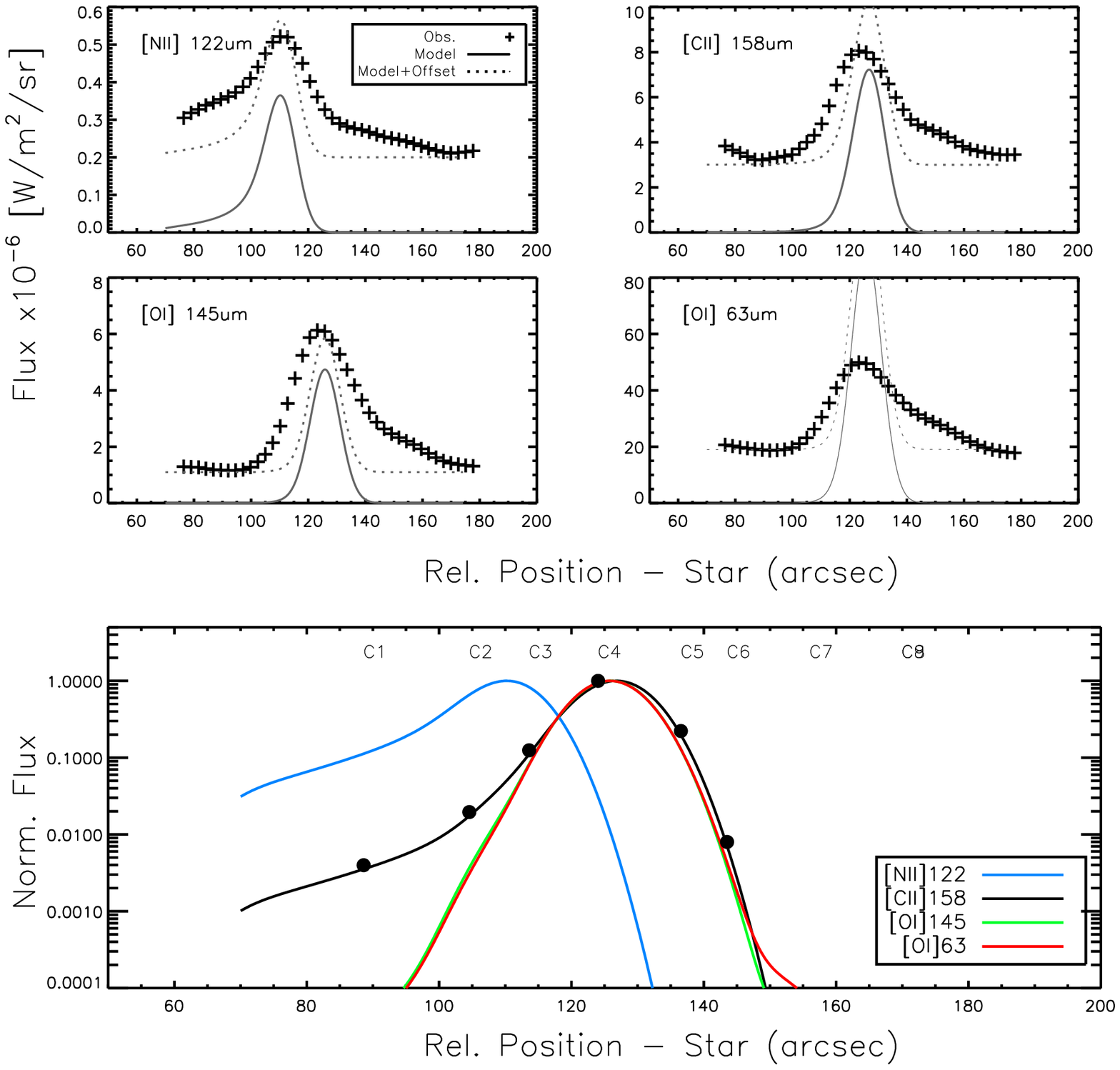}
  \end{center}
  \caption{Observed (plus symbols) and modelled (solid and dotted
    lines) profiles along the cut shown in Figure~4 (see Sect. 5)
    measured with respect to the distance to the exciting star
    $\theta$$^1$ Ori C.  Observed and modelled fluxes are convolved to the
      158$\mu$m PSF, where the dotted line includes an offset due to
    background emission (see text). The first four panels compare for each line
    the observed convolved profiles with a Cloudy model, where this
    has been scaled to the peak of the \ion{[O}{I]} (145 $\mu$m)
    emission. The bottom panel compares the normalized model
    profiles. In this panel, the solid dots represent the position of
    the reference points in the adopted cut (Fig. 4a).}
\end{figure}

We have modelled the line emission across the Bar using the radiative
transfer code Cloudy \citep{fer98}. This code computes the chemistry
and radiative transfer at the surface of a molecular cloud (in our
case assumed to be a plane-parallel semi-infinite slab) illuminated by
FUV photons from the ionising star(s).  We adopt a Kurucz model for
the star at 39\,600~K \citep{pel09}, and ISM abundances
\citep{sav96,mey97,mey98}. In order to reproduce the spatial
stratification we adjust the starting point of the HII region and the
density. Namely, the ionised region is adjusted to start at 0.134~pc
(67$\arcsec$) from the star. The
adopted gas density is set to a constant value of 3200~cm$^{-3}$ \citep{pel09} in
the ionised region, and is then coupled with a profile density for the
PDR as described in \citet{ara11}, with a density scaled to 6$\times$10$^4$
cm$^{-3}$ at the peak of the \ion{[O}{I]}  and \ion{[C}{II]} and emission
(projected distance of 0.246~pc).  No tilting is assumed for the Bar and the depth of
  the PDR  along the line of sight is adjusted to 0.35~pc.  

To illustrate the comparison between the observed fluxes and the model
(convolved to the 158$\mu$m PSF)
we assume the cut illustrated in Figure~4a. This cut is also adopted
for the discussion in our accompanying papers by \citet{hab11} and
\citet{ara11}. It is made to pass through the exciting star $\theta^1$
Ori C (5h 35m 16.46s, -5d 23m 23.17s) and an arbitrary point in the
Bar (5h 35m 21.82s, -5d 24m 59.18s). The chosen point in the Bar
minimizes effects of other contaminating stars in the direction of the
cut that can affect the dust \citep{ara11}, and allows us to sample different
conditions in the region. In Figure~6 the solid line
  represents the output of the model, and the dotted line the model
  plus an offset (see point 3.) From the comparison in Figure~6 we
find that:

\begin{enumerate}

\item This simple model does a very good job in reproducing the
  observed spatial stratification of the peak positions of the
  \ion{[N}{II]} line at 111.5$\arcsec$, and the \ion{[C}{II]} and
  \ion{[O}{I]} lines at 123.5$\arcsec$ (top and middle panels). This
  stratification is the result of the attenuation of the incident
  radiation field across the Bar and is sensitive to the density
  profile.

\item The model also shows a gradual increase of the \ion{[N}{II]}
  line as we approach the Bar (better seen in the bottom panel) which
  is also seen in the observations, while in contrast the oxygen lines
  show a more abrupt increase.  This is the result of the ionisation
  structure. The model also shows a gradual increase for the
  \ion{[C}{II]} line but this is not clear from the observations. This
  could be the result of it being more easily excited (low temperature
  and density) compared to the oxygen lines and/or contribution of the
  \ion{[C}{II]} line from the ionised phase.

\item It is striking that the observed profiles do not fall to zero as
  the model (solid lines) does both before and after the Bar. This
  background emission is significant, amounting to about $\sim$18\% of
  the peak emission for the \ion{[O}{I]} 145$\mu$m line and $\sim$38\%
  for the other three lines.   These offsets have been added to
    the model emission as the dotted lines in the Figure.  This
    suggests the presence of another PDR(s), probably from the
    background cavity in the molecular cloud, which must produce the
    base of emission that we measure.  This PDR is probably not
  entirely face-on because the strong intensity observed, which may
  result from a limb brightening effect, cannot be reproduced by a
  face-on PDR model.

\item The model line profiles are narrower than the observations. This
  could be due to the Bar being tilted to the observer ($\lesssim$10$\degr$), an effect
  already inferred in other studies \citep[e.g.,][]{pel09}. 
    Alternatively, a highly structured medium would allow FUV
    radiation to permeate the region and heat gas on larger spatial
    scales.

\item In order to reproduce the absolute flux, and assuming that the
  base emission from the additional PDR(s) has an additive effect on
  top of the emission of the Bar (dotted line), we need to assume a
  depth of the PDR in the model of 0.3~pc. The depth given above is
  larger than, but comparable to, the width of the Bar ($\sim$0.06~pc) as
  expected for a tilted edge-on Bar.

\item In this model we take into account optical depth effects across
    the PDR\footnote{We assume both thermal and micro-turbulence line
      width with a turbulence velocity of 3~km~s$^{-1}$.} but not
    along the line of sight. However, we can see that the model (with
    offset) reproduces the peak of emission of the \ion{[N}{II]}
    122$\mu$m, \ion{[O}{I]} 145$\mu$m fairly well, and the
    \ion{[C}{II]} 158$\mu$m within factor 1.2. This indicates that, except for the \ion{[O}{I]}
    63$\mu$m line, the emission of these lines are not dominated by optical effects.

\end{enumerate}

A more detailed modelling of the region including additional PDR/s
(even the effects of clumpiness and tilting) and dust emission is outside
our scope here but will be presented in a future paper. Still, it is
encouraging to see how a simple model (single edge-on PDR) can be used
to at least reproduce with a fair degree of success the observations
of such a complex environment.
 



\section{[CII] from the PDR and HII region}

In the ISM, the \ion{[C}{II]}158$\mu$m line is important in the
  study of the cooling and chemistry of PDRs. In extragalactic studies it is
  important for redshift determinations, and for the extent to which
  its luminosity is a measure of the star formation rate
  \citep[SFR,][]{sta91, mei07,luh03}. With a low ionisation potential
  of 11.3~eV, this line can originate both in the PDR and in the HII
  region. It is thus important to characterise in detail its
  contribution in different environments. Because of the high
  radiation field and high density in the Orion nebula, one expects
  that most of the \ion{[C}{II]} should originate from the PDR.  The
  PACS maps show in great detail the distribution of the \ion{[C}{II]}
  and \ion{[N}{II]} lines where we can investigate this issue.

\begin{figure}
  \begin{center}
   \includegraphics[width=7.0cm,angle=90]{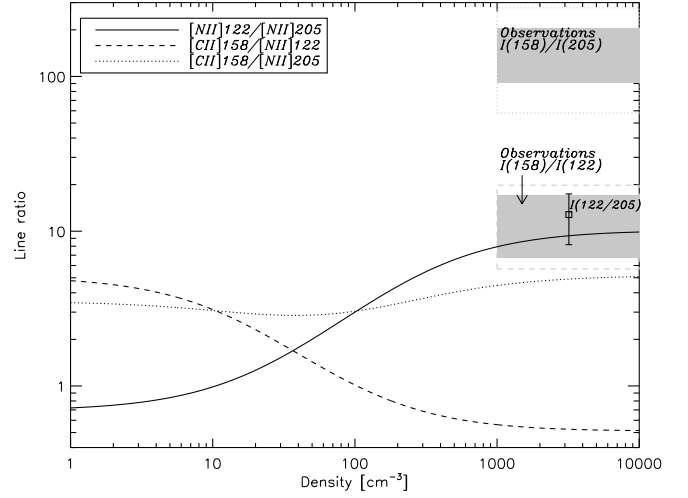}
  \end{center}
  \caption{Line intensity ratios for different combinations of the \ion{[N}{II]} (122 and 205$\mu$m) and \ion{[C}{II]}
158$\mu$m lines. Theoretical curves are shown by the solid, dotted, and
dashed lines. Grey areas indicate the range of observed ratios, and
the enclosing boxes represent the range of uncertainties in the ratios. The square point is the ratio of the \ion{[N}{II]} 122/205 lines.}
\end{figure}

We have therefore quantified the contribution of \ion{[C}{II]}
158$\mu$m and \ion{[N}{II]} 122$\mu$m to the Bar as traced by the
\ion{[C}{II]} emission (Bar(CII)), and as traced by the \ion{[N}{II]}
emission (Bar(NII)). These are represented as areas enclosed by the
solid and dashed lines, respectively, in Figure~4a. This combined region, which
we name Bar(CII$+$NII), encompasses the peak emission of the PDR and
ionised region.  We find that 76\% of the
\ion{[C}{II]} emission is coming from the PDR region (Bar(CII))
relative to the total emission in the Bar(CII+NII). The remaining 31\%
comes from the Bar(NII) region\footnote{These quantities do not add to
  100\% because there is an overlapping region as it is explained in
  Sect. 3.}. This 31\% contribution to the Bar(NII) region could include
emission from the ionised region or from the
additional PDR (see Sect. 5). To estimate this, and ignoring
 confusion from  tilting effects, we can use the close
relation of the \ion{[C}{II]} line with the \ion{[O}{I]}145$\mu$m
line.  For a broad range of densities between 10$^3$ and 10$^5$
(typical of PDRs), the \ion{[C}{II]}/\ion{[O}{I]}145$\mu$m ratio
varies between 0.5 and 2 \citep{fer98,kau99,lep06}. The observed
\ion{[O}{I]}145$\mu$m in the Bar(NII) is about 2$\times$10$^{-6}$
W~m$^{-2}$~sr$^{-1}$, from which we predict a PDR \ion{[C}{II]}
emission between 1 to 4$\times$10$^{-6}$ W~m$^{-2}$~sr$^{-1}$. We can
then compare this range to our observed \ion{[C}{II]} emission of
4$\times$10$^{-6}$ W~m$^{-2}$~sr$^{-1}$ in this region.  Thus, from
the original 31\% of \ion{[C}{II]} in this region, we deduce that
either all could come from the additional PDR, or the HII region could
contribute to at most 24\%.

Following the analysis by \citet{obe11} in the Carina nebula, we
  have compared the expected \ion{[C}{II]}/\ion{[N}{II]} ratio in the
  ionised medium with the observed values in the Bar(NII) region. This
  can be used to get an additional estimate of the contribution from
  the ionised region to the \ion{[C}{II]} line. To do this we have
  complemented our observations with the SPIRE-FTS data of the
  \ion{[N}{II]}205$\mu$m line \citep{hab11}, and have calculated the
  theoretical curves for the
  \ion{[N}{II]}122$\mu$m/\ion{[N}{II]}205$\mu$m,
  \ion{[C}{II]}158$\mu$m/\ion{[N}{II]}205$\mu$m, and
  \ion{[C}{II]}158$\mu$m/\ion{[N}{II]}122$\mu$m ratios at a
  temperature of 9000~K \citep{nie11}. The \ion{[C}{II]}/\ion{[N}{II]}
  line ratios depend on the relative ionic abundance of CII and
  NII. We use the carbon and nitrogen abundances adopted in Sect.~5,
  coupled with the ionisation fraction of \ion{C}{II} and \ion{N}{II}
  for the HII conditions in the Orion nebula as given by
  \citet{rub85}, to derive a relative ionic abundance of
  \ion{[C}{II}/\ion{N}{II]}$\sim$1.6. These theoretical curves are
  shown in Figure 7\footnote{The \ion{[C}{II]}158$\mu$m/\ion{[N}{II]}
    (122, 205$\mu$m) curves can be applied to any other object by
    simply scaling them to the ratio between the relative abundance of
    \ion{[C}{II}/\ion{N}{II]} in the object and that assumed here
    (1.6).}. The \ion{[N}{II]}122$\mu$m/\ion{[N}{II]}205$\mu$m ratio
  can be used to determine the electron density. The observed ratio
  (square in the figure) falls in the non-linear regime of the
  theoretical curve (solid line) but can serve to obtain an upper
  limit in the density of 1000 cm$^{-3}$, consistent with our adopted
  value of 3200 cm$^{-3}$ \citep{pel09}. The observed
  \ion{[C}{II]}158$\mu$m/\ion{[N}{II]}205$\mu$m, and
  \ion{[C}{II]}158$\mu$m/\ion{[N}{II]}122$\mu$m line ratios, grey
  areas in the figure, are clearly above the theoretical curves
  (dotted and dashed lines respectively). Taking into account the
  uncertainties in the line fluxes (enclosing boxes in the figure),
  the observed ratios indicate that no more than 9\% of the
  \ion{[C}{II]}158$\mu$m comes from the ionised region. Allowing
  for an uncertainty of about a factor 2 in the relative abundance of
  \ion{C}{II}/\ion{N}{II} and collisional strengths used to calculate
  the theoretical curves, we can safely
  conclude that in the Bar less than 18\% of the \ion{[C}{II]}
  emission originates in the ionised medium. This is a slightly
  lower threshold than our previous estimate.

\citet{abe05} made some modifications to the Cloudy code to, among
others things, estimate the contribution of several PDR lines from the H\,II
region. They apply their model to the starburst galaxy NGC\,253 where
they find that about 30\% of the \ion{[C}{II]} comes from the ionised
medium. More recently, \citet{moo11} using PACS data have mapped the
FIR emission of the \ion{[C}{II]} 158$\mu$m line in spiral galaxy
M33. They find that between 20 to 30\% of this emission comes from the
ionised medium. Their values  are slightly higher than
our determinations, which is not surprising since we are probing dense PDR in the Orion Bar while in these galaxies lower density HII regions could dominate the emission. We note, however, that this comparison is hampered
by the difference between the larger scales (where different
regions are mixed) that these studies probe, and the small scales
probed in our study.

\section{Cooling}

In order to establish the role of the \ion{[C}{II]} and \ion{[O}{I]}
lines in the cooling of the PDR we have compared their contribution to
that of other relevant cooling lines in the mid- and FIR. We have
included in this calculation the H$_2$ rotational lines from the
ISO/SWS, and $^{12}CO$ rotational lines (J=4-3 to J=21-20), $^{13}CO$
(J=5-4 to J=14-13), H$_2$O, and CH$^+$ from the PACS and SPIRE
instruments \citep{hab10,job11} . To make this comparison we use the
same position (5h35m21s, -5\degr~25\arcmin~18\arcsec) of the H$_2$
measurements by ISO to derive the contribution of each line (or
cascade of lines). This position is indicated by the white square in
  Figure~4a. For this comparison we convolve all the data to the
  largest beam size (SPIRE  40\arcsec). We find that in this region the \ion{[C}{II]} and
\ion{[O}{I]} lines studied in this paper contribute 90\% of the total
power emitted by all these lines, with the \ion{[O}{I]} 63$\mu$m line
contributing 72\% of this emission. The CO, H$_2$, and CH$^+$
contribute, respectively, 5, 4, and less than 1\%. These estimates
accentuates the importance of the \ion{[O}{I]} 63$\mu$m line to the
cooling budget from the gas lines in these regions.  This fact is also highlighted by the
  relative strength of this line compared to the other lines in the
  region (see Fig. 6).

\section{Summary and Conclusions}

We have presented the first HSO observations of the
\ion{[C}{II]}158$\mu$m, \ion{[O}{I]}63$\mu$m and 145$\mu$m, and
\ion{[N}{II]}122$\mu$m lines of the Orion Bar. Its angular resolution has
allowed us to map the spatial distribution of these lines in
unprecedented detail.

The \ion{[C}{II]} and \ion{[O}{I]} maps peak at the same position and
fall close to the peak emission of PAHs (as traced by the Spitzer/IRAC
8$\mu$m band). The \ion{[N}{II]} peaks slightly closer to $\theta$$^1$
Ori C with a small region of overlap with respect to the other PDR
lines.  Within the Bar we can distinguish between knots (clumps) of
emission, about 0.01-0.02~pc in size (6 to 10\arcsec), and
  which are 16\% higher in flux than the interclump medium. These
  clumps could  be photo-evaporated by the FUV
  inside the PDR. These knots of emission are
better seen in the \ion{[O}{I]}63$\mu$m map as this line offers the
best resolution (having the smallest PSF), it is also seen in the
\ion{[O}{I]}145$\mu$m and \ion{[C}{II]}158$\mu$m maps.

The \ion{[C}{II]}158$\mu$m correlates very well with the \ion{[O}{I]}
145$\mu$m emission. The \ion{[C}{II]}158$\mu$m line does not correlate
with the \ion{[N}{II]} in the ionised region as some studies have
suggested.  The combined information on the \ion{[N}{II]} and
\ion{[O}{I]} lines provides a great diagnostic to estimate the
emission from the \ion{[C}{II]}158$\mu$m line, and to distinguish
between an origin in an ionised or neutral region.  The ratio between
the \ion{[O}{I]} 145$\mu$m/63$\mu$m lines show the effect of the
opacities, where the \ion{[O}{I]} 63$\mu$m line becomes self-absorbed
at high column densities.


We have modelled the emission of the lines with the photo-ionisation
code Cloudy and reproduce the relative position of the lines.
The emission profiles reveal that, in addition to the Bar,
there is a significant background emission all over the region
(present for all four lines). This points to the presence of
additional PDR(s). This should be follow up with detail radiative
transfer models to infer the physical conditions and coupled to what
has been learned from the emission of dust.

The 
\ion{[C}{II]} line can come from the neutral and ionised medium.  We have made different estimations of its contribution and find that  most of the [CII] emission originates in the PDR(s) ($>$82\%).  Using ancillary ISO and Herschel data we have
calculated the total power emitted by the most relevant cooling lines
from the mid- to FIR (\ion{[O}{I]}, \ion{[C}{II]}, CO, H$_2$O,
CH$^+$).  We show that the power emitted by the atomic \ion{[C}{II]}
158$\mu$m, and \ion{[O}{I]} 63$\mu$m and 145$\mu$m lines account for
90\% of the power emitted in the region by all of the cooling lines
considered, with the \ion{[O}{I]} 63$\mu$m line contributing 72\% of
the total. This emphasizes the predominant role of the latter in the
cooling process from emission lines of these regions.


\begin{acknowledgements}

  We would like to thank the referee, David Hollenbach for his comments and suggestions. JBS wishes to acknowledge the support from a
  Marie Curie Intra-European Fellowship within the 7th European
  Community Framework Program under project number 272820.  HCSS,
  HSpot, and HIPE are joint developments by the Herschel Science
  Ground Segment Consortium, consisting of ESA, the NASA Herschel
  Science Center, and the HIFI, PACS and SPIRE consortia. JBS thanks
  E. Romano-Diaz for help with contour manipulation. We thank
  V. Lebouteiller for the use of the PACSman software and discussion
  on data reduction.

\end{acknowledgements}




\end{document}